# Surface acoustic wave modes in two-dimensional shallow void inclusion phononic crystals on GaAs

Edward Muzar and James A. H. Stotz

Department of Physics, Engineering Physics & Astronomy,

Queen's University, Kingston, Ontario, Canada

## Abstract

Surface acoustic waves in two-dimensional phononic crystals consisting of a square array of shallow, two to three micron deep cylindrical void inclusions are studied computationally via the finite element method. For the [110] propagation direction on a (001) GaAs half-space, the conventional Rayleigh wave modes, the layered substrate-associated Sezawa and Lamb modes, the high frequency longitudinal surface waves and bulk waves exhibit hybridization and modal interaction. The longitudinal, vertical shear and horizontal shear bulk wave dispersions are observed to be significant thresholds for surface acoustic waves on a shallow phononic crystal. This results in dramatic changes in the attenuation and surface boundedness properties that enable supersonic modes with diminished attenuation in otherwise largely bulk wave-radiative dispersion branch continua of modes.

## Introduction

Phononic crystal (PnC) structures for controlling surface acoustic wave (SAW) propagation have been studied as an alternative to traditional metal gratings for applications requiring SAW guiding, reflecting and electronic device integration [1]. One significant design challenge of PnCs for SAW devices is the bulk mode conversion losses from leaky SAWs [2]. Brillouin zone folding, due to an inhomogeneous substrate periodicity, can induce even a subsonic SAW into Pseudo-SAW/Leaky-SAW (PSAW) behavior, which radiates bulk wave energy; such processes are referred to as Brekhovskikh attenuation [3]. Pseudo-Rayleigh wave propagation along directions neighboring the [110]-direction of a semi-infinite (001) GaAs homogeneous substrate are intrinsically radiative resulting from a phase velocity that is greater than the slowest bulk wave and a coupling to bulk quasi-shear waves with horizontal polarization [4]. However, Rayleigh waves travelling exactly along the [110] crystal direction are ordinary SAWs rather than PSAWs [5] with no bulk wave coupling necessary [4] to satisfy both the equation of motion for the anisotropic medium and the traction-free boundary conditions due to the semi-infinite substrate [6]. For cubic crystals such as GaAs (001), bulk waves exactly along [110] behave as pure modes: longitudinal, shear vertical and shear horizontal [7] and hence couple/interact with SAWs in a simpler manner where bulk motions are either entirely sagittal or out-of-plane. Such bulk mode conversion loss does not decrease per inclusion with larger inclusion arrays but rather converges to some finite loss amount per inclusion [8]. Thus, for large array PnC structures, such bulk radiative losses usually scale linearly with array size and must be accounted into the design of practical devices including guided SAW devices that may require long propagation lengths.

Early works attempted to understand experimental results by calculating the attenuation properties of certain surface gratings on an isotropic material in the Brillouin zone below and above the two bulk dispersion lines [9], [10]. One approach minimized bulk wave conversion of SAWs on GaAs by optimizing a grating (i.e. one-dimensional PnC) orientation with respect to Rayleigh waves along [110] on (001) GaAs for an oblique SAW reflector [11] while in another study, a one-dimensional grating composed of aluminum on a glass substrate achieved SAW hybridization with longitudinal bulk waves with a resulting bandgap [12]. A thin film with periodic modulation of elastic properties was also shown to achieve Rayleigh and Sezawa wave mode hybridization, where negligible bulk wave conversion was possible for a localized mode in an otherwise radiative SAW mode branch within the supersonic region of the bandstructure [13]. In addition to the small zone boundary bandgap, such a hybridization of SAW modes facilitates the emergence of modes with vanishing attenuation and a large bandgap that is not located at a zone boundary due to the folded retrograde Rayleigh wave and unfolded prograde Sezawa wave superposition (or vice versa) forming approximate standing waves [3], [14]. Such modes on an elastic layered structure with low attenuation can be termed "embedded or bound states in the radiation continuum" (BICs) [15] or an "isolated true surface wave" (ITSW) [16]. These hybridization-based gaps, in addition to Bragg gaps and scatterer resonant gaps, constitute one of three types of band gaps [17]. Adding another component to the complexity of the model, the transmission in an anisotropic substrate has subsequently been studied via the finite element method (FEM) for bulk to SAW conversion [18] and vice versa [19] in a finite PnC, and further theoretical studies include a one-dimensional PnC/grating via eigenfrequency and transmission spectra that showed various SAW modes [20].

Prior to modern PnCs, there was experimental study of SAW devices with void-based grating reflectors, first by one-dimensional grooves [21], and eventually by two-dimensional arrays on a substrate that showed grating-like behavior [22]. The research study in [22] considered the two-dimensional array consisting of a thick metallic overlay as periodic only along the SAW propagation direction, with irregular spacing laterally, in order to suppress coherent scattering effects. Theoretical work has also focused on the scattering by "dots" (inclusions) [23]. Further information of these prior developments are found in [24]. Following the successes in photonic crystals, early work describing a modern PnC, consisting of a square lattice SAW with infinitely deep cylindrical AlAs inclusions that are arranged in a periodic, two-dimensional square lattice and embedded in GaAs, demonstrated bandgap formation for SAWs and PSAWs [25]. It was later determined a SAW bandgap can also occur for shallow cylindrical inclusions [26]. In these previous two studies, the acoustic impedance mismatch was low resulting in lesser effects on the band structure. SAW attenuation at bandgap frequencies in deep cylindrical air inclusion PnCs was studied for an isotropic substrate via experimental transmission spectra and compared to calculated band structures [27]. A two-dimensional shallow cylindrical void inclusion PnC for intrinsically subsonic SAWs was later studied via eigenfrequency and frequency domain FEM, which showed more prominent influences from the PnC on transmission. The wavevectors were considered in the complex domain at the zone boundary, and the interior substrate boundary condition was the Dirichlet type with zero displacement [28] and also with a shallow PnC layer composed of separate material with voids [29]. Experimental and simulation of wet-etched non-cylindrical shallow void inclusion, square lattice PnCs were also studied [30], [31].

For modelling acoustic excitations in a semi-infinite domain, identifying and characterizing SAW modes from bulk-like acoustic modes must be considered, and it has been addressed by various similar methods in gratings and PnC systems where modes of both types exist [14], [32], [33], [34], [20], [35]. For identifying and characterizing SAW modes, the surface boundedness and attenuation will be studied. We will characterize the surface boundedness of SAW modes by the dimensionless strain energy ratio $E_{\text{surface}}/E_{\text{domain}}$ of the integrated strain energy contained in the surface half of the computational domain, $E_{\text{surface}}$, to the integrated whole computational domain, $E_{\text{domain}}$. Thus, the strain energy ratio is a parameter characterizing the strain energy distribution and can vary from 0 to 1, where 0 indicates no strain energy is contained in the surface half, 1 indicates all strain energy is in the surface half, and a uniform distribution of energy is characterized by 0.5. In this paper, the dimensionless strain energy ratio criterion $E_{\text{surface}}/E_{\text{domain}}$ will be greater than 0.6 in the band structure diagrams for enhanced contrast and for clearer understanding of SAW properties in the PnCs. For comparison, the ratio of the imaginary and real components of mode frequency, $\text{Im}(\omega)/\text{Re}(\omega)$ will also be used to define the SAW attenuation as in [14].

In addition to the aforementioned strain energy ratio to characterize SAW boundedness, an additional ratio characterizing the sagittal behavior of the modes is utilized. Since Love waves and other horizontally polarized displacement modes are physically possible in addition to the sagittally polarized Rayleigh and Sezawa modes [36], it is necessary to discern the polarization of modes. On a semi-infinite GaAs (001) substrate, the Rayleigh wave along [110], the *x*-direction, has displacement components [37] such that $u_x = f_x(x,z) \neq 0$, $u_z = f_z(x,z) \neq 0$ and $u_y = 0$ where z is along $[001]$ and y is along $[1\bar{1}0]$. For this present study, a dimensionless ratio parameter is utilized, the sagittal displacement polarization ratio of a mode, *r*, defined as,

$$r = \frac{\int (u_x^* u_x + u_z^* u_z) \mathrm{d}V}{\int (u_x^* u_x + u_y^* u_y + u_z^* u_z) \mathrm{d}V}$$

where $u_x$, $u_y$ and $u_z$ are displacement vector ***u*** components, and d*V* is a differential volume element to the integral that is over the whole computational domain. This parameter is analogous to the ratio of transverse polarization introduced for the selection of Love modes found in [35]. The integrand terms $u_i^* u_j \delta_{ij}$ are squared displacement component magnitudes for selection favoring SAWs with concentrated, relatively large displacements, which is expected at the surface for SAWs. A ratio of 0 indicates the displacement vector is normal to the sagittal plane and a ratio of 1 indicates the displacement vector is entirely in the sagittal plane. For this work, the *r* minimal threshold will be set at 0.6 for consideration in the results for the selection of modes with significantly sagittal behavior.

The mechanical behavior of substrates with thin films is also important to understanding the nature of SAWs present as the shallow-holes of the PnC can be viewed as a structured metamaterial-like layer on top of a semi-infinite substrate. SAW propagation behaviors have been categorized for flat, isotropic, homogeneous, uniform thin films on semi-infinite isotropic substrates based on the bulk shear phase velocity difference $\Delta v = v_{thin\ film} - v_{substrate}$ of the thin film and substrate [38] [39]. A Sezawa or Lamb wave can propagate for $\Delta v < 0$ and a Stoneley wave can propagate for $\Delta v \cong 0$ [39]. Although such wave propagation categories exist for layered isotropic materials, the more intricate behavior of SAW propagation for anisotropic layered materials is less clear, where a natural anisotropy is due to crystal structure and an artificial anisotropy is due to PnC/gratings on the surface.

In this current study, a PnC composed of a GaAs matrix with cylindrical void inclusions, and hence a high impedance mismatch, is explored for its propagation and attenuation properties via eigenfrequency analysis. The FEM is used to model and simulate the shallow cylindrical inclusions forming the PnC. FEM is suitable for studying SAWs in periodic structures and subsequent attenuation properties [40]. We also restrict the attention to certain SAWs with predominantly *xz* sagittal plane motion, where *x* and *z* are along [110] and [001] crystal directions of a (001) half space in this case. As such, the sagittal planes refer to those parallel to the surface normal and to Γ-X direction of the PnC. The Γ-X and Γ-M directions of the square PnC lattice are parallel with the GaAs [110] and GaAs [100] directions, respectively. The PnC plane is located at the GaAs-vacuum surface interface with a finite, uniform thickness. The PnC plane normal vector is parallel with the GaAs (001) plane vector normal to the surface. The depths of the inclusions to be discussed include 2, 2.5 and 3 µm; hence, all three thicknesses can be conceptualized as a surface array [24] (i.e. grating) in a computational domain thickness or depth of 50 µm. The deeply penetrating nature of bulk waves is sensitive to the finite domain depth that is utilized. Therefore, some parameters that are sensitive to the boundary of the computational domain (such as the imaginary frequencies) may have deviations in the quoted values resulting from the finiteness of the model.

## Simulation

FEM modelling and computations were performed in COMSOL Multiphysics [41], which has also been used to study SAWs on shallow hole PnCs [28]. An eigenfrequency study was performed on a three-dimensional linear elastic anisotropic domain for real frequencies up to 500 MHz. A homogeneous half-space of (001)-cut GaAs was truncated in the model to a 50 µm deep computational domain. The cylindrical inclusions had a surface filling fraction of 0.564 as in [25] and uniform array depths of 2, 2.5 or 3 µm. Additionally, a substrate without any PnC structure was also simulated as a control. The PnC lattice parameter was 4 µm for a square lattice with Γ-X oriented along [110] of (001) GaAs. Meshing of the 2 µm deep inclusion PnC computational domain included 14448 tetrahedral elements. The GaAs elastic stiffness constants and density are values used previously by Tanaka and Tamura [25]. The surface boundary that included the cylindrical inclusion surface was specified as a free boundary condition; hence, it was tractionless for the GaAs-void interface. The side boundary conditions of the unit cells were specified as Bloch-Floquet for a set of reduced wavevectors $q$ (= $k/k_X$) ranging from 0.5 to 1 with a step size of 0.002, where $k$ is the wavevector magnitude in the Γ-X direction and $k_X$ is the wavevector magnitude at the X point. A previous study used a fixed boundary at the lower truncated interface of the domain to model phononic crystals for SAWs [42]; however, we implement an absorbing boundary condition (ABC) at the truncated domain limit in the substrate as a dissipation mechanism into the bulk substrate for bulk waves in order to determine the non-zero imaginary frequency component of the eigenmodes. With the real and imaginary parts of frequency from such damping, the attenuation characterizing the SAW attenuation could then be determined. A similar ABC was also selected in a previous study of SAW losses to bulk waves in interdigitated transducers/ metal electrode gratings on lithium tantalite [43]. The plotted longitudinal, vertical shear and horizontal shear bulk dispersions are based on the same stiffness and density from the simulation, and using the bulk wave slowness formulae in Auld [7].

## Results and Discussion

Figure 1 (a-f) shows the bandstructures in the ΓX-direction of a square phononic crystal with the ΓX-direction oriented along the [110]-direction of (001)-cut GaAs for 2 µm (subplots a & b), 2.5 µm (subplots c & d) and 3 µm (subplots e & f) deep inclusions; the color-coded markers display: (subplots a, c, e) logarithm of reciprocal attenuation, and (subplots b, d, f) the strain energy ratio $E_{\text{surface}}/E_{\text{domain}}$. The attenuation defined as $\text{Im}(\omega)/\text{Re}(\omega)$ [14] is exhibited as a logarithmic reciprocal attenuation scale and hence, it is $-\log_{10}(\text{Im}(\omega)/\text{Re}(\omega))$ so that the correspondence with the strain energy ratio is clearer. In each plot, the reduced wavevector is $q$ $(=\frac{k}{k_X})$, where $k$ is the wavevector, $k_X$ $(=\frac{\pi}{a})$ is the Brillouin zone X point wavevector with $a$ as the PnC lattice parameter equaling 4 µm. The uncertainty in the real frequency in the simulations with respect to the simulation model is estimated at order of 0.01 MHz since the substrate-only simulation with two degenerate solutions at the zone boundary simulated as 356.09 and 356.11 MHz, a difference indicative of the FEM approximation and meshing. The lowest frequency Rayleigh dispersion branch has a perturbed dispersion behavior with lower phase velocity compared to an ordinary one on a semi-infinite homogeneous substrate, as is expected for a substrate with a slow surface layer [39]. At a reduced wavevector of 1, the zone boundary, this lowest branch has a frequency of 315.55 MHz for 2 µm deep inclusions, compared to the unperturbed case at a frequency of 356.10 MHz based on the separate control simulation of a substrate without a PnC. With the deeper inclusions, the Rayleigh wave unfolded band is progressively lowered in frequency to the extent that, for reduced wavevectors near or at the zone boundary, the Rayleigh wave is below all three bulk wave dispersion lines and, thus, is in fact a subsonic perturbed Rayleigh wave in that interval.

A finer scale of the two parameters characterizing the attenuation and surface boundedness of the modes for the unfolded Rayleigh mode (1st branch) are shown in Figure 2 (a & b). At the zone boundary ($q$ = 1), the Rayleigh wave attenuation is greatest for 2 µm deep inclusions, and progressively lessens with 2.5 and 3 µm deep inclusions. The 3 µm deep inclusion PnC even exhibits a lower attenuation than the substrate since the dispersion is significantly below the horizontal shear line. At $q$ = 0.5, the attenuations are progressively larger with increasing inclusion depth given the attenuation behavior is dictated for this larger wavelength by surface roughness arising from larger scattering cross-sections with the 3 µm geometry now having an attenuation over an order of magnitude larger than the substrate. In the interval $0.5 < q < 1$, the attenuation of the first SAW branch for 3 µm deep inclusions has a dramatic variation in the attenuation of almost five orders of magnitude resulting from crossing the horizontal shear bulk line and thus traversing the supersonic to subsonic barrier. The small fluctuations of the strain energy ratio and attenuation ratio at intermediate $q$ occur due to mode crossings with horizontal shear horizontally polarized bulk-like waves that only weakly interact with the sagittal polarized SAWs. The 315.55 MHz mode at $q = 1$ is pictured in Figure 3 (a) showing the modal profile of a Rayleigh wave.

At the Brillouin zone boundary for the 2 µm deep inclusions in Figure 1 (a & b), the next higher frequency mode is at 318.94 MHz and is separated by a bandgap of about 3 MHz above the lowest mode. The modal displacement profile is seen in Figure 3 (b) and is seen to be a quarter-wavelength out of phase with the mode at 315.55 MHz seen in Figure 3 (a) and thus confirming that these two modes at the zone boundary define the Rayleigh mode bandgap. This bandgap is relatively small compared to the band edge frequencies. Generally, this can be expected for non-hybrid, zone boundary bandgaps [14], and to a lesser extent due to the a solid matrix with less dense solid inclusions [44], and similar results have been observed for SAWs in PnCs with both deep but low elastic contrast [25] and shallow inclusions with high elastic contrast void inclusions [28]. The dispersion shows negative group velocity down to a reduced wavevector of about 0.83, a threshold at which the modes of smaller wavevector are seen to have a hybridized behavior illustrated in Figure 3 (g). The modes at $q < 0.83$ take on a more Sezawa mode profile with lower $q$, rather than a Rayleigh wave mode profile seen at the zone boundary, as indicated by the reversing of the direction of surface motion [45]. At the same time, the dispersion also becomes more bulk-like and follows the shear vertical bulk line as has been previously shown [46] [38]. The hybridized Rayleigh-Sezawa behavior of the dispersion for the zone folded Rayleigh that is located at 318.94 MHz at the zone boundary is similar to the behavior seen in [3], where the zone folded Rayleigh wave hybridization with the Sezawa wave forms a maximum far from the Brillouin zone boundary. Figure 2 (c - d) provides a direct plot of the attenuation and surface boundedness for the second mode and demonstrate that each of the Rayleigh modes in a PnC are highly coupled to modes associated with the shear vertical line, since the attenuation increases and surface boundedness decreases close to or above the shear vertical line. In the substrate control, these horizontal shear and vertical shear bulk waves artificially interact with the SAW due to the domain folding effect and contribute to the small variations in attenuation and strain energy ratio, which allows for a glimpse of a minute bulk mode interaction behavior in the limit of a zero elastic contrast for this FEM simulation.

Although the attenuation is low, this deep Sezawa wave does not confer additional attenuation lowering properties in this PnC with 2 μm deep inclusions possibly due to the already low attenuation of the folded Rayleigh wave, and hence a negligible bulk radiation source contribution to balance with the Sezawa wave bulk wave radiation. The Sezawa wave contribution is seen to become very bulk-like for $q$ < 0.83 and is seen in Figure 3 (g) for $q$ = 0.83 (compare to the mode profile of $q$ = 0.82 in Appendix Figure 1 (c) to see more pronounced Sezawa wave behavior). The bulk-like behavior is characteristic of higher order Rayleigh waves like Sezawa waves as these wave dispersions asymptotically approach the bulk shear line [47]. Thus, in this current system, the Sezawa wave bulk radiation component is not balanced by the Rayleigh wave bulk radiation component and hence, this particular hybridized SAW does not produce a BIC. The Rayleigh mode does exist past the shear vertical line but is seen to be attenuated in Figure 1 (a & b), which is verified by the bulk wave radiation seen in the mode profiles of Figure 3 (f). These two modes are on discontinuous segments of the folded Rayleigh wave and constitute two dispersion segments of an avoided crossing. The sagittally polarized surface/bulk-like modes whose dispersions would intersect the Rayleigh dispersion results in the hybridization and the avoided crossing. There are two additional avoided crossing hybridizations at lower wavevectors including one at $q$ = 0.8, and another close to $q$ = 0.5 that is not entirely visible due to below-threshold strain energy ratio values for those particular modes. There is also a crossing with a modified dispersion in the neighborhood of the intersection with a deep sagittal mode near $q$ = 0.76. The crossing and anti-crossing behavior is a feature of the extent of damping due to the coupling with these sagittal polarized waves [48]. Overall, the folded Rayleigh wave above the shear vertical line has a dispersion that is curved with a mostly positive group velocity. This is caused by a shallower penetration depth, which in turn situates the SAW even more in the slower surface layer containing the voids, causing a progressively lowered phase velocity at lower $q$ for the folded Rayleigh mode. The attenuation constant dramatically increases for the dispersion curves above the shear vertical sound line but the SAW behavior is better characterized here with the strain energy ratio that has more scale contrast. The energy ratio shows a well surface bounded wave near the dispersion maximum for $q$ = 0.7 to 0.8. Thus, there may be viable BICs possible for in such branches above the vertical shear line with optimized PnC designs.

The third dispersion branch of the PnC with 2 μm deep inclusions appears as an unfolded Sezawa mode branch. Its mode profile at the zone boundary with real frequency 406.02 MHz is seen in Figure 3 (c) and shows a large amplitude component deep into the substrate. This third branch runs parallel with the Sezawa mode of the second branch at low reduced wavevectors, with both dispersions following the shear vertical bulk line; this indicates both SAWs have strong coupling with the shear vertical bulk modes. Although both Sezawa waves are coupled to the shear vertical modes, the two PSAW dispersions remain slightly separate. Below a reduced wavevector of 0.98, the Sezawa wave is no longer well confined to the surface; it is plotted partially at lower wavevectors according to the strain energy threshold of 0.6 to exhibit the degeneracy with a shear vertical bulk wave. The unconstrained modes with large penetration depths has also been seen elsewhere, including [20]. This is an indicator that the shear vertical bulk-like waves readily hybridize with these SAWs and consequently increase the penetration depth of the PSAWs.

Just above the third dispersion unfolded Sezawa branch the zone boundary (406.02 MHz) is a fourth dispersion branch with a single mode that is immediately below the shear vertical line in Figure 1 (a & b). It is a mode from the folded dispersion of the third dispersion branch of the Sezawa wave. This dispersion branch consists of many small segments in proximity to the folded shear vertical line and behaves like a cascade of dispersion plateaus. This behavior is due to multiple Lamb waves of sagittal polarization due to the computational domain interacting with the folded Sezawa mode and the shear vertical bulk modes (not shown). These sagittal waves are shear vertical bulk waves and Lamb waves of the effective thin surface layer due to the PnC. Thus, this dispersion branch indicates PSAW dispersions that are close to bulk wave dispersions can be highly modified by the modal interactions and segmented with very little continuity. The exact behavior of computational domain Lamb modes is highly dependent on the domain depth, and hence, the folded Sezawa dispersion is affected by the computational domain—specifically, the finite depth of the substrate.

The fifth dispersion branch frequency for the PnC with 2 µm inclusions is also plotted in Figure 1 (a & b) along with the mode displacement profiles for $q$ = 0.71, 0.85, and 1 in Figure 3 (e) (h) and (d), respectively. The fifth dispersion branch is separated into five separate dispersion segments due to the hybridization with shear vertical bulk-like modes and also the folded third branch. Overall, the fifth branch appears to have a flat dispersion of a PSAW wave mode associated with a local surface resonance [33] and the PSAW resemblance is verified by the modal profiles of the $u_z$ sagittal component (Figure 3). Above $q$ > 0.75, the PSAW has more similar modal profiles to typical Rayleigh waves. The branch has discontinuous segments due to hybridizations with the folded shear vertical mode (shown) and Lamb modes (not shown) that are deep and typically lossy. Away from these shear modes, the behavior at lower wavevectors ($q < 0.73$) exhibit different PSAW properties. Between the folded shear vertical line and unfolded longitudinal line, the attenuation has a minimum at reduced wave vectors near 0.71, for which the modal displacement profiles is presented in Figure 3 (e) and shows a deep longitudinal ($u_x$) behavior. In comparison, at a reduced wavevector of 0.85 on this same dispersion segment (Figure 3 (h)), the mode structure is considerably more Rayleigh wave-like in behavior. At lower wavevectors, the fifth dispersion branch degenerates to a longitudinal bulk-like wave, of which a segment is plotted.

To further characterize this interesting dispersion segment of the fifth branch between the longitudinal and shear vertical modes, the reciprocal attenuation and strain energy ratio are included in Figure 2 (e). At larger $q$, past the attenuation minimum near $q = 0.71$, this dispersion segment is closer to the folded shear vertical bulk wave line and has a general trend towards larger attenuation and lower surface boundedness (Appendix Figure 1 (a) shows a more bulk longitudinal wave at $q = 0.68$ to compare). Within this trend of higher attenuation with larger wavevector, there is also the folded Sezawa mode situated in a hybridized complex with a bulk shear vertical mode along this approach towards larger $q$. This results in a peak of lowered attenuation and higher strain energy ratio. This may be indicative that the radiation of bulk waves from the reversed direction of the folded Sezawa destructively interferes with the radiation of bulk waves from the forward travelling Rayleigh wave. This is similar to the lowering of attenuation mechanism as reported in other studies [3], [14]. Interestingly, the strain energy ratio peak near $q = 0.85$ occurs at slightly lower $q$ than the reciprocal attenuation peak; hence, a particular balance of hybrid contributions from a deep Sezawa wave and vertical shear hybrid wave with the more superficial SAW achieves a lower attenuation with a larger contribution from the deep Sezawa and vertical shear hybrid wave. The hybrid mode is pictured in Figure 3 (h) for $q = 0.85$ (Appendix Figure 1 (b) has a non-hybrid mode profile at $q = 0.80$ to compare on the same branch and Appendix Figure 1 (d) has a more Sezawa mode profile at $q = 0.86$ on the same branch). Based on the reciprocal attenuation peak and surface boundedness, this Sezawa and Rayleigh hybrid exhibits BIC behavior. The attenuation lowering for the SAW near the bulk longitudinal wave line at reduced wavevector of 0.71 is explained [49] in that the wave at that point can be described as a high frequency surface longitudinal wave. This SAW mode therefore has a deeper longitudinal component and is located between the shear and longitudinal bulk sound lines [50], as is the case in the neighborhood of the low attenuation point through hybridization similar to [14], where SAW hybridization lowers attenuation, albeit through Rayleigh-Sezawa wave hybridization in that case. This particular mode is described as a SAW that is composed of a Rayleigh wave hybridized with a bulk longitudinal resonance [12]. Thus, for the current system, it is corroborated that a single material, in this case GaAs with a two-dimensional PnC surface, can produce non-symmetry protected BICs without a two-material uniformly or a periodically patterned thin film that is consists of a separate material [15], [51].

In addition to the aforementioned modes, the two bandstructures for a PnC with 2.5 μm and 3 μm deep inclusions in Figure 1 (c – f) show an additional PSAW mode below the shear vertical bulk wave dispersion. This is expected and explained as a Lamb wave associated with the surface perturbation of the thin film [52]; a Lamb wave of higher order than the Sezawa wave associated with the thicker effective slow surface layer. The flat dispersion of fifth branch containing the longitudinal resonance SAW is also seen following the longitudinal bulk line but lacking the hybridization with the shear vertical bulk-like modes at larger reduced wavevectors near the folded shear vertical bulk line. The low attenuation interval does not extend as far as in the 2 μm deep inclusion case due to this lack of additional hybridization with the Sezawa and shear vertical bulk-like mode hybridization. This lack of an extended low attenuation interval for this mode indicates that the additional hybridization from the shear vertical bulk like with the Sezawa mode is providing destructive interference for the bulk radiation from the longitudinal resonance for the 2 μm deep inclusion PnC, in a manner similar to previous work [14]. The folded Sezawa dispersion is largely lost in the 2.5 μm deep inclusion case, but is more robust with lower attenuation in the 3 μm case due to more favorable hybridization for SAW behavior with Lamb modes. The hybrid mode BIC does appear to occur with the flat dispersion Rayleigh and the folded Sezawa in each of these cases.

The sound line concept [53] for surface guided modes in PnC structures may be extended to account for PSAW dispersion situated BICs with low attenuation between longitudinal and shear bulk phase velocities, instead of only subsonic SAWs. The low attenuation may be acceptable for certain PSAW-PnC device designs and also allow more design options for systems where subsonic SAWs are usually of primary importance. The BIC concept in elasticity by Maznev and Every [14] [15] is supported and extended with findings in this present work for two-dimensional surface PnCs. The existence of low attenuation longitudinal resonance modes and Rayleigh-Sezawa hybrid SAWs in a GaAs PnC structure exemplifies that BICs in such a system are a possibility. Through a surface topographic pattern on GaAs, the semiconductor can permit the manifestation of such features as the non-symmetrically protected longitudinal resonance, and the Rayleigh-Sezawa wave hybrid but also effectively slow surface subsonic Rayleigh waves in a crystal direction that is otherwise supersonic without local PnC defects.

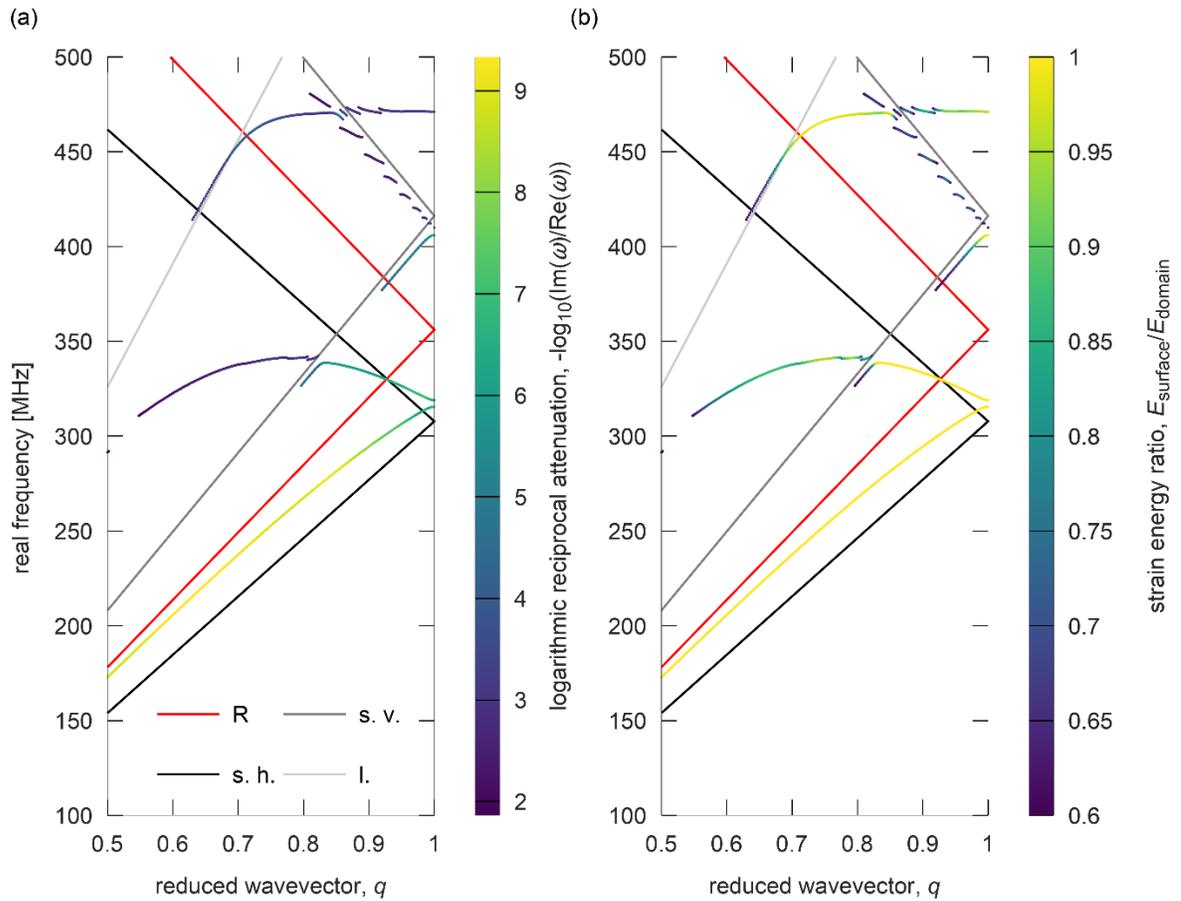

Figure 1 (a & b): The band structure in the ΓX-direction of the PnC oriented along the [110] crystalline direction of (001) GaAs for 2 μm deep cylindrical inclusions. The bandstructure is shown with color-coded markers of (a) logarithm of reciprocal attenuation and (b) strain energy ratio. The Rayleigh wave (R) and the three bulk waves, shear horizontal (s. h.), shear vertical (s. v.), and longitudinal (l.), are shown as zone folded lines.

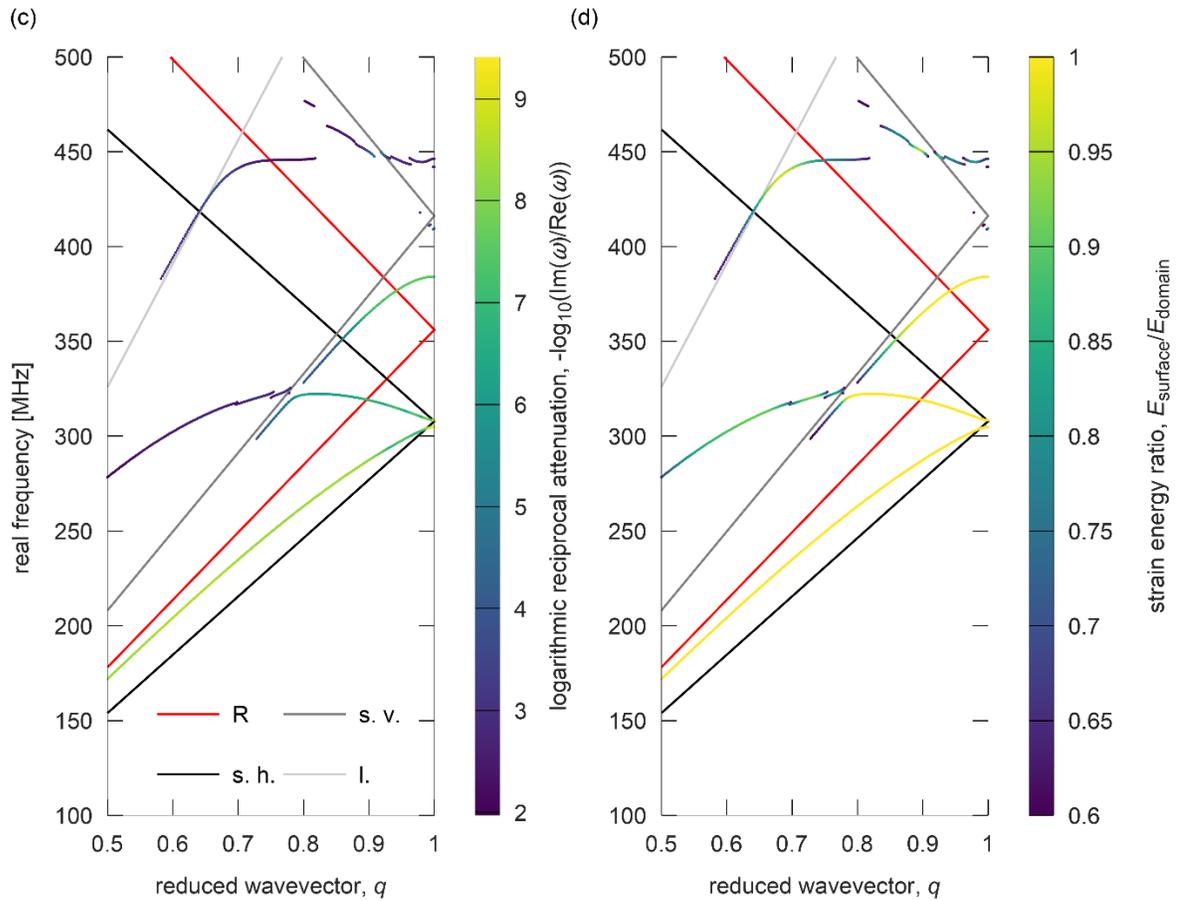

Figure 1 (c & d): The band structure in the ΓX direction of the PnC oriented along the [110] crystalline direction of (001) GaAs for 2.5 μm deep cylindrical inclusions. The bandstructure is shown with color-coded markers of (a) logarithm of reciprocal attenuation and (b) strain energy ratio. The Rayleigh wave (R) and the three bulk waves, shear horizontal (s. h.), shear vertical (s. v.), and longitudinal (l.), are shown as zone folded lines.

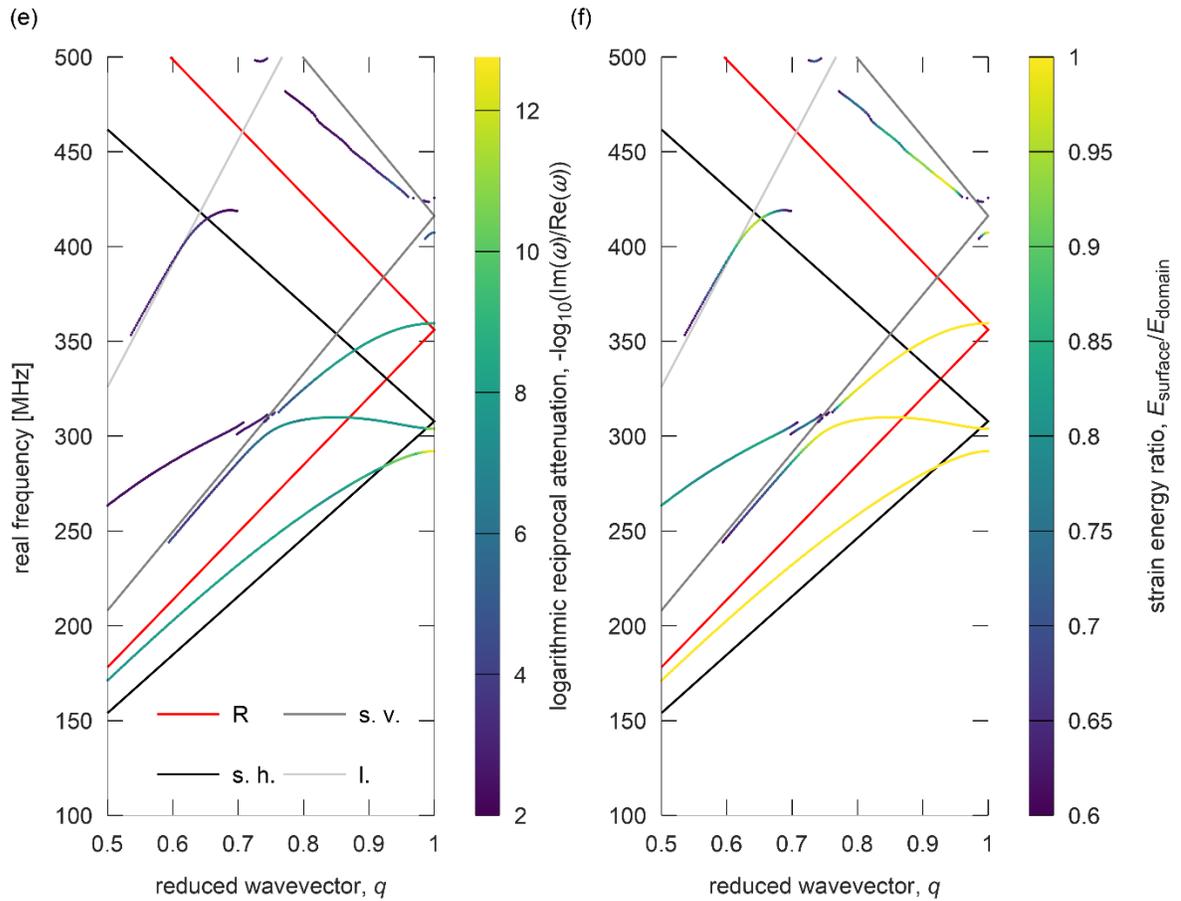

Figure 1 (e & f): The band structure in the ΓX direction of the PnC oriented along the [110] crystalline direction of (001) GaAs for 3 μm deep cylindrical inclusions. The bandstructure is shown with color-coded markers of (a) logarithm of reciprocal attenuation and (b) strain energy ratio. The Rayleigh wave (R) and the three bulk waves, shear horizontal (s. h.), shear vertical (s. v.), and longitudinal (l.), are shown as zone folded lines.

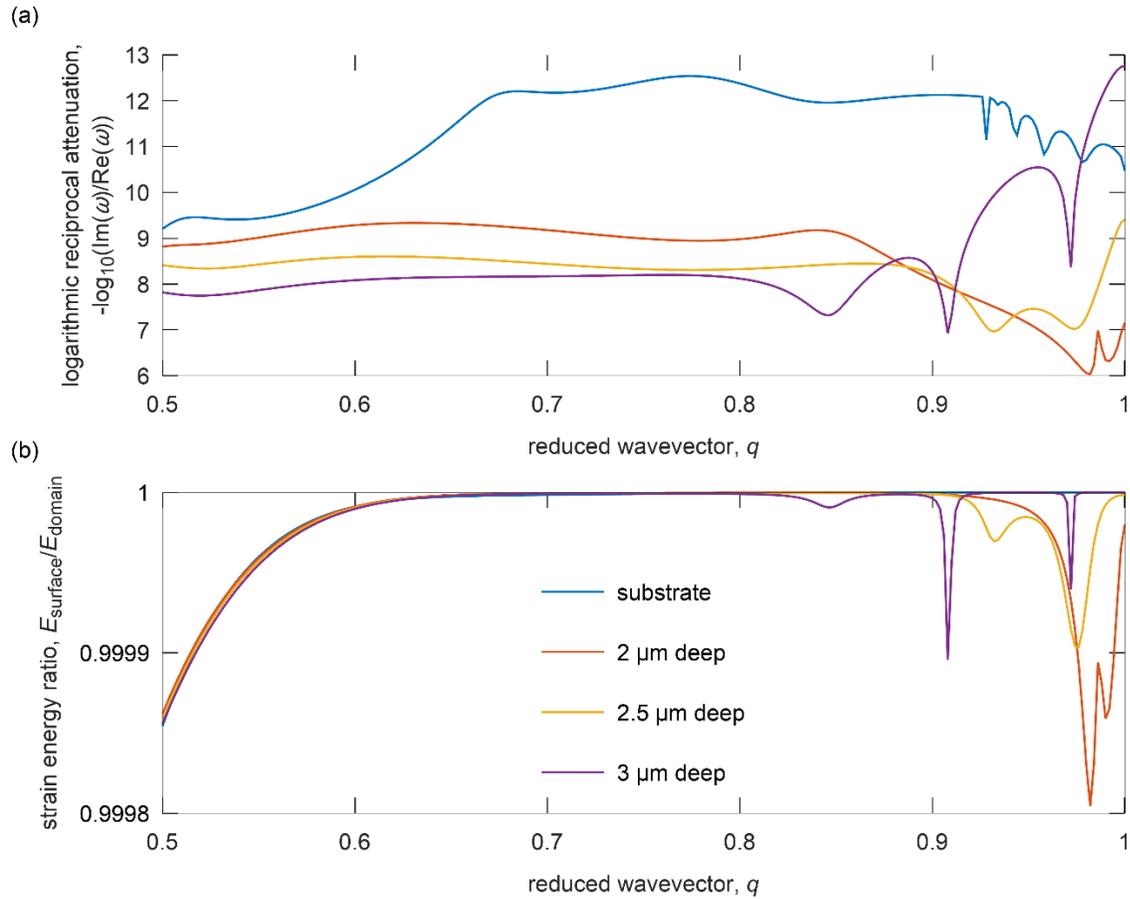

Figure 2 (a & b): The substrate, and perturbed, 2, 2.5 and 3 µm deep inclusion PnC, unfolded Rayleigh wave (a) logarithmic reciprocal attenuation and (b) strain energy ratio compared to the ordinary Rayleigh wave contained in the first Brillouin zone. The key is presented in (b). The dispersion is seen in Figure 1 (a & b) as the lowest frequency dispersion curve with the modal frequency of 315.55 MHz at the zone boundary.

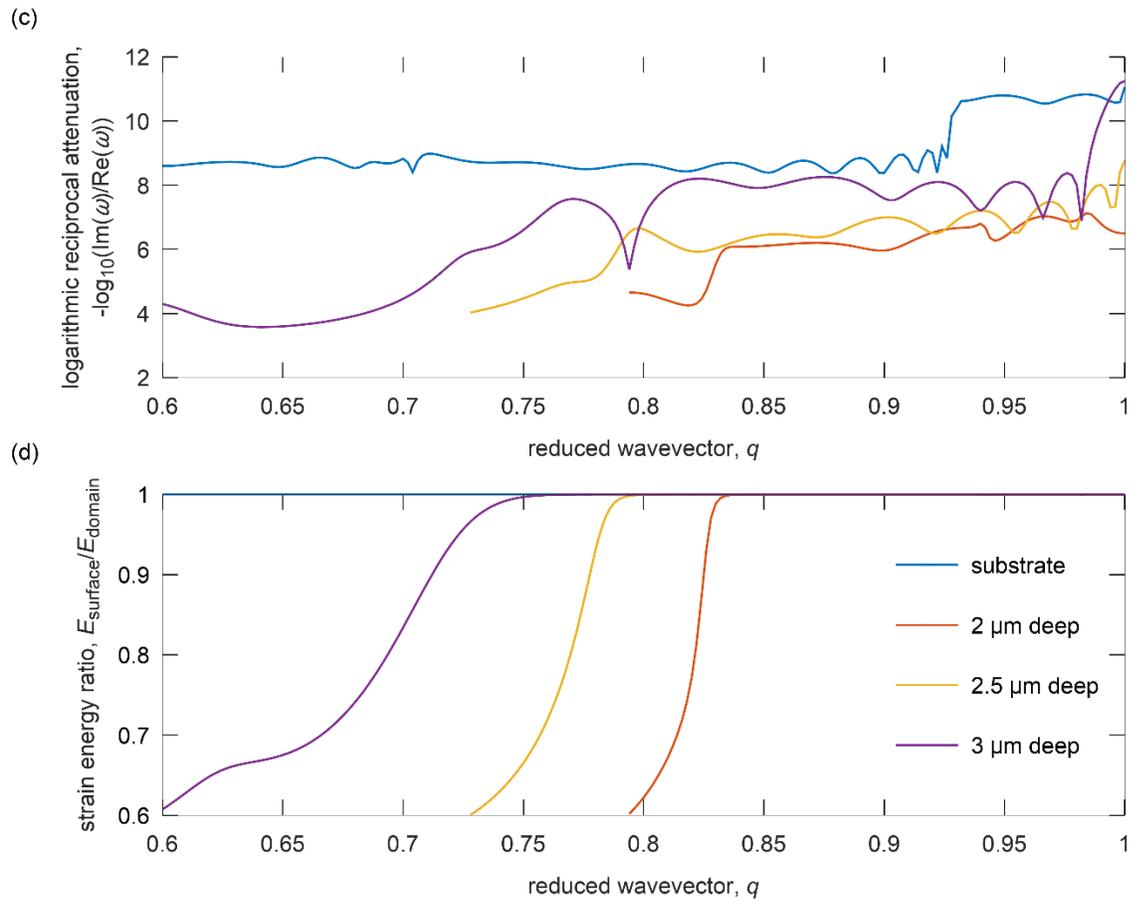

Figure 2 (c & d): The substrate, and perturbed, 2, 2.5 and 3 µm deep inclusion PnC, zone folded ordinary Rayleigh wave logarithmic reciprocal attenuation and strain energy ratio, plotted versus reduced wavevector. The attenuation is presented on the left side y-axis as logarithmic reciprocal attenuation and the right side y-axis as strain energy ratio with corresponding colors to the curves. The dispersion is seen in Figure 1 (a & b) as the second lowest frequency dispersion curve and is only the segment containing the mode frequency of 318.94 MHz at the zone boundary, and is below the shear vertical (s. v.) bulk line, but above the unfolded shear horizontal (s. h.) bulk line.

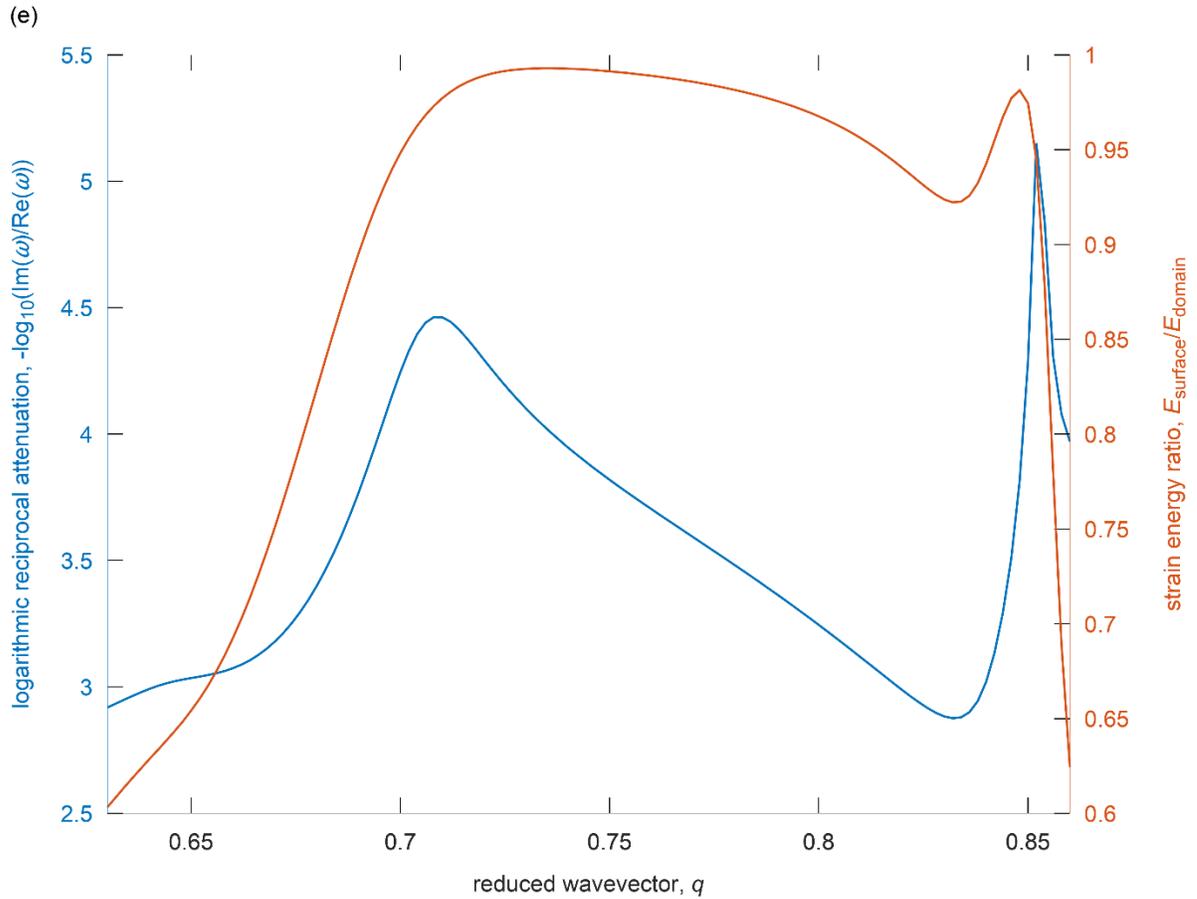

Figure 2 (e): The 2 µm deep inclusion PnC longitudinal resonance SAW (fifth branch) logarithmic reciprocal attenuation and strain energy ratio. The attenuation is presented on the left side y-axis as logarithmic reciprocal attenuation and the right side y-axis as strain energy ratio with corresponding colors to the respective curves. The dispersion is seen in Figure 1 (a & b) as the dispersion curve adjacent to the bulk longitudinal dispersion line.

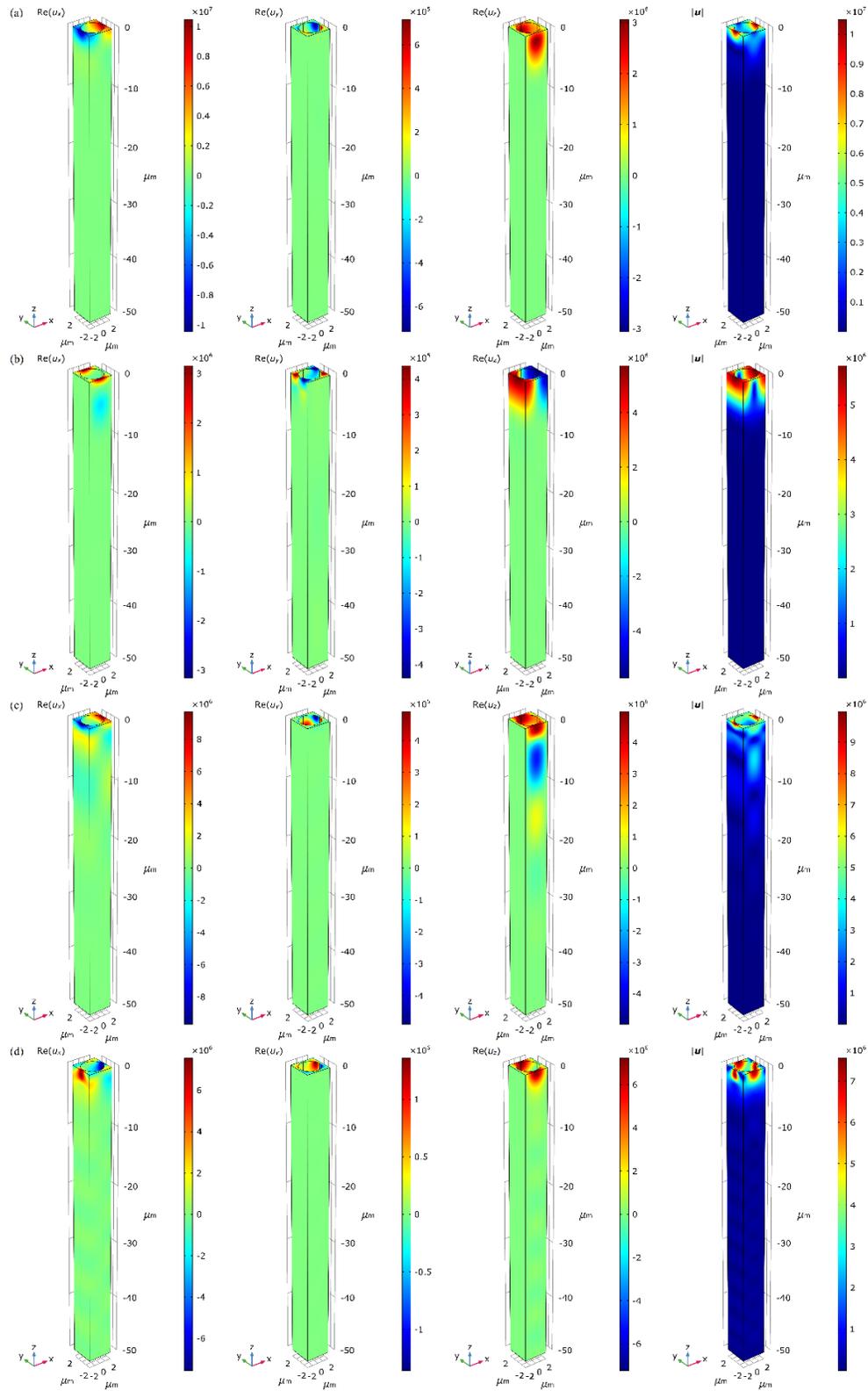

Figure 3 (a - d): The 2 μm deep inclusion PnC displacement field components (real parts) and its magnitude for modes at the X point, the zone boundary with q = 1, at real frequencies (a) 315.55 MHz, (b) 318.94 MHz, (c) 406.02 MHz, and (d) 471.09 MHz.

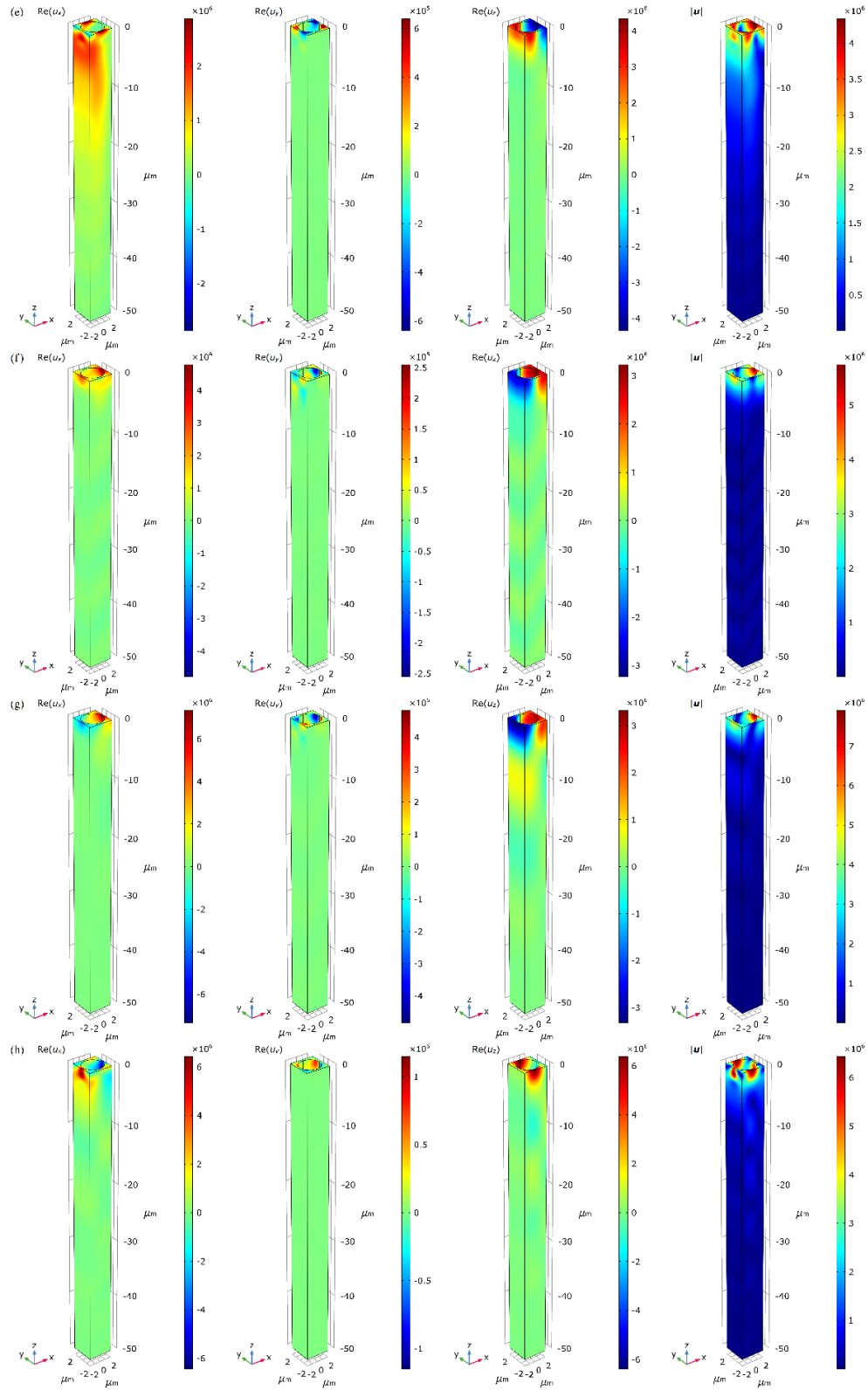

Figure 3 (e - h): The displacement field components (real parts) and its magnitude for modes at (e) q = 0.71; real frequency 457.38 MHz, (f) q = 0.75; real frequency 340.7 MHz, (g) q = 0.83; real frequency 338.51 MHz, and (h) q = 0.85; real frequency 469.85 MHz.

## Conclusion

We have performed a thorough mapping of surface acoustic modes for waves propagating along the ΓX-direction of a two-dimensional square lattice of voids on a GaAs surface via FEM simulation. The presence of the PnC significantly increases the ability of surface and bulk modes to hybridize with the result often being a scattering of strain energy into bulk modes. We have characterized SAW modes using both a strain energy ratio that dictates the energy near the surface as well as through an attenuation parameter resulting from an absorbing boundary condition on the simulation domain's semi-infinite truncation boundary at a surface depth of 50 μm. Through this analysis, SAW modes and instances of BIC behavior for shallow-inclusion, two-dimensional PnCs revealed interesting attenuation properties as was predicted generally for supported thin-films and may be useful in applications requiring low bulk wave losses or Brekhovskikh attenuation such as guided wave devices. This is particularly important for SAWs propagating in the [110] direction including Rayleigh waves that are intrinsically supersonic.

## Acknowledgements

The authors would like to gratefully acknowledge the following: the open source GNU Octave (V4) project, whose software was used to generate two-dimensional plots in this work [54]; CMC Microsystems for the provision of products and services that facilitated this research, including Design Tools and computational resources and support; Computations were also performed on resources and with support provided by the Centre for Advanced Computing (CAC) at Queen's University in Kingston, Ontario. The CAC is funded by: the Canada Foundation for Innovation, the Government of Ontario, and Queen's University; Natural Science and Engineering Research Council of Canada for supporting this work under Discovery Grant RGPIN/05701-2014.

# Appendix

Appendix Figure 1 contains modal displacement profiles of four modes associated with modes pictured in Figure 3. The modes are as follows: Appendix Figure 1 (a) is on the fifth dispersion and exemplifies the behavior of the modes following the longitudinal line; (b) is on fifth dispersion and is relatively lossy, far from hybridization effects; (c) is on the second dispersion showing a hybrid markedly more Sezawa wave motion; (d) is on the fifth dispersion and is an image of a hybrid mode with more Sezawa wave-like character.

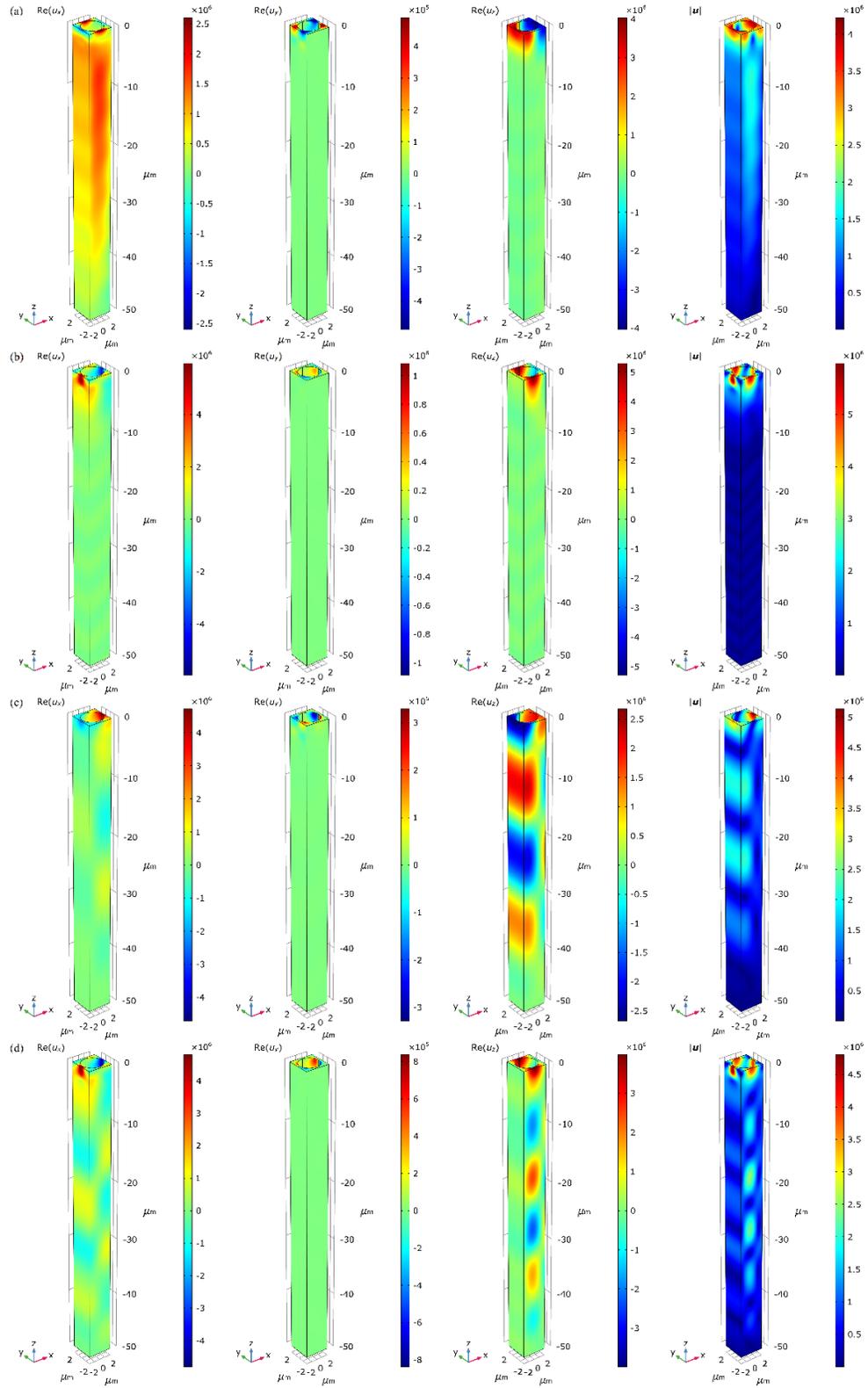

Appendix Figure 1 (a – d): The 2 μm deep inclusion PnC displacement field components and its magnitude for modes at (a) q = 0.68; real frequency 444.09 MHz, (b) q = 0.80; real frequency 469.76 MHz, (c) q = 0.82; real frequency 336.15 MHz, and (d) q = 0.86; real frequency 467.08 MHz.